\documentclass[10pt]{article}
\usepackage{amsmath}
\usepackage{amsfonts}
\usepackage{amssymb}
\usepackage{amsthm}
\usepackage{amssymb}
\usepackage{hyperref}
\usepackage{url}
\usepackage{indentfirst}
\usepackage{tikz}
\usepackage{booktabs}
\usetikzlibrary{shapes.geometric}
\usepackage{filecontents}
\usepackage{booktabs}
\usepackage[noend]{algpseudocode}
\usepackage[margin=1.5in]{geometry}  % For reducing margin
\usepackage{algorithm}
\usepackage{graphicx}
\usetikzlibrary{shapes,positioning}
\usepackage[utf8]{inputenc}
\usepackage{hyperref}
\usepackage{caption}
 
\usepackage[justification=centering]{caption} % Figures caption
\usepackage{graphicx}
\usepackage{caption, subcaption}
\usepackage{amsthm}

\usepackage{tikz}
\usetikzlibrary{arrows}
\usetikzlibrary{positioning}

\newtheorem{theorem}{Theorem}

\newtheorem{proposition}{Proposition}

\newtheorem{remark}{Remark}

\def\cO{\mathcal{O}}

\def\P{\mathcal{P}}

\def\ZZ{\mathbb{Z}}
\def\ZZ1{\mathbf{Z}_{\geq 1}}

\def\RR{\mathbb{R}}

\title{On Nash-Solvability of Finite Two-Person 
Tight Vector Game Forms}

\author{Vladimir Gurvich\\
vgurvich@hse.ru and vladimir.gurvich@gmail.com\\
RUTCOR, Rutgers University, Piscataway, NJ, United States;\\
National Research University Higher School of Economics, Moscow, Russia
\and
Mariya Naumova\\
mnaumova@business.rutgers.edu\\
Rutgers Business School, Rutgers University, Piscataway, NJ, United States\\
}

\begin{document}

\maketitle

\begin{abstract} 
We consider finite two-person normal form games.
The following four properties of  
their game forms are equivalent:
(i) Nash-solvability, (ii) zero-sum-solvability, (iii) win-lose-solvability, and 
(iv) tightness. 
For  (ii, iii, iv) this was shown by Edmonds and Fulkerson in 1970. 
Then, in 1975, (i) was added to this list and it was also shown that 
these results cannot be generalized for $n$-person case with  $n > 2$. 
\newline 
In 1990, tightness was extended 
to vector game forms ($v$-forms) and 
it was shown that such  $v$-tightness and  
zero-sum-solvability are still equivalent, 
yet, do not imply Nash-solvability. 
These results are applicable to several classes 
of stochastic games with perfect information.
\newline 
Here we suggest one more extension of tightness 
introducing $v^+$-tight vector game forms ($v^+$-forms).  
We show that such $v^+$-tightness and Nash-solvability are equivalent 
in case of weakly rectangular game forms and positive cost functions. 
This result allows us to reduce 
the so-called bi-shortest path conjecture 
to $v^+$-tightness of $v^+$-forms. 
However, both (equivalent) statements remain open.  
\end{abstract}

\section{Nash-solvability of tight game forms} 
\subsection{Game forms and cost functions}
We consider finite normal form games of two players, Alice and Bob. 

% \begin{remark} 
% We will mention $n$-person games in the last two sections, 
% but all games will be assumed finite in this paper. 
% \end{remark} 

A {\em game form} is a mapping  $g : X \times Y \rightarrow O$, 
where  $O$  is a set of possible outcomes, while 
$X$ and  $Y$  are the sets of strategies of Alice and of Bob, respectively. 
These three sets are finite.
%$O$  % = \{o_1, \ldots, o_m\}$  is the set of outcomes and  
% $X = \{x_1, \ldots, x_m\}$, $Y = \{y_1, \ldots, y_n\}$ 

\begin{remark}
In this paper we restrict ourselves and the players 
by their pure strategies. The mixed ones are not mentioned.
\end{remark}

Furthermore, let 
$r^A : O \rightarrow \RR$  and  $r^B : O \rightarrow \RR$  
be {\em rewards} or {\em payoffs} of the players, 
that is, if outcome  $o \in O$  appears, Alice and Bob 
get  $r^A(o)$  and  $r^B(o)$, respectively. 
Both players are maximizers. 
The triplet  $(g, r^A, r^B)$  is 
called a {\em finite two-person normal form game}, 
or just a game, for short. 

\begin{remark}
``Separating"  rewards from game forms 
allows us to make the latter responsible 
for the structural properties of games, 
which hold for arbitrary rewards.
\end{remark}

\subsection{Nash equilibria and Nash-solvability}
A pair of strategies  $(x,y) \in X \times Y$  
is called a {\em strategy profile} or a {\em situation}.

\medskip 

A situation $(x,y)$  is called a {\em  Nash equilibrium}  (NE) if 

\medskip 

$r^A(g(x,y)) \geq r^A(g(x',y))$  for any  $x' \in X$  and 
$r^B(g(x,y)) \geq r^B(g(x,y'))$  for any  $y' \in Y$; 

\medskip 
\noindent 
in other words, if no player can profit by replacing her/his strategy, provided the opponent keeps his/her strategy unchanged. 
Another equivalent reformulation is as follows:

\smallskip 

$(x,y)$ is a NE if and only if $y$ is a best response for  $x$  and $x$ is a best response for  $y$. 

\smallskip 

This concept was introduced by John Nash in \cite{Nas50,Nas51}. 

\medskip 

A game form  $g$  is called {\em Nash-solvable} 
(NS) if the corresponding game  $(g, r^A, r^B)$  
has a NE for any rewards  $r^A$  and  $r^B$.   
In particular, $g$  is called 
{\em zero-sum-solvable}  
(respectively, {\em win-lose-solvable}) 
if game  $(g,r^A,r^B)$  has a saddle point   
for any $r^A$  and  $r^B$  such that  
$r^A + r^B \equiv 0$ 
(respectively, if  $r^A + r^B \equiv 0$  and 
$r^A, r^B$  take only values $\pm 1$).
In the zero-sum case we assume that 
Alice is a maximizer, while Bob is the minimizer.

\medskip 

Obviously, NS implies zero-sum-solvability, which in its turn implies win-lose-solvability.  

\subsection{Examples of game forms}
Several examples are given in Figure \ref{f1}, where
game forms are represented by tables with rows, columns, and 
entries labelled by $x \in X$, $y \in Y$, and  $o \in O$.  
Mapping  $g$  is assumed to be surjective, 
but not necessarily injective,
that is, an outcome $o \in O$ 
may occupy an arbitrary array in the table of  $g$.

\begin{figure*}[h]
%\centering
\captionsetup[subtable]{position = below}
\captionsetup[table]{position=top}
\hspace*{4em}
\begin{subtable}{0.3\linewidth}
\centering
\begin{tabular}{|c|c|}
\hline
$o_1$ & $o_1$ \\ \hline
$o_2$ & $o_3$ \\ \hline
\end{tabular}
\caption*{$g_1$}               \label{tab:g1}
\end{subtable}
\hspace*{-3em}
\begin{subtable}{0.3\linewidth}
\centering
\begin{tabular}{|c|c|c|c|}
\hline
$o_1$ & $o_1$ & $o_2$ & $o_2$ \\ \hline
$o_3$ & $o_4$ & $o_3$ & $o_4$ \\
\hline
\end{tabular}
\caption*{$g_2$}
\label{tab:g2}
\end{subtable}
\hspace*{-2.5em}
\begin{subtable}{0.3\linewidth}
\centering
\begin{tabular}{|c|c|c|}
\hline
$o_1$ & $o_1$ & $o_3$\\ \hline
$o_1$ & $o_2$ & $o_2$\\ \hline
$o_3$ & $o_2$ & $o_3$ \\ \hline
\end{tabular}
\caption*{$g_3$}               \label{tab:g3}
\end{subtable}

\vspace{0.5cm}

\hspace*{4em}
\begin{subtable}{0.3\linewidth}
\centering
\begin{tabular}{|c|c|c|}
\hline
$o_1$ & $o_1$ & $o_3$\\ \hline
$o_1$ & $o_1$ & $o_2$\\ \hline
$o_4$ & $o_2$ & $o_2$ \\ \hline
\end{tabular}
\caption*{$g_4$}
\label{tab:g4}
\end{subtable}
\quad
\hspace*{-4em}
\begin{subtable}{0.3\linewidth}
\centering
\begin{tabular}{|c|c|c|c|}
\hline
$o_1$ & $o_2$ & $o_1$ & $o_2$\\ \hline
$o_3$ & $o_4$ & $o_4$ & $o_3$\\ \hline
$o_1$ & $o_4$ & $o_1$ & $o_5$\\ \hline
$o_3$ & $o_2$ & $o_6$ & $o_2$ \\ \hline
\end{tabular}
\caption*{$g_5$}
\label{tab:g5}
\end{subtable}
\quad
\hspace*{-3.5em}
\begin{subtable}{0.3\linewidth}
\centering
\begin{tabular}{|c|c|}
\hline
$o_1$ & $o_1$ \\ \hline
$o_1$ & $o_2$ \\ \hline
\end{tabular}
\caption*{$g_6$}
\label{tab:g6}
\end{subtable}

\vspace{0.5cm}

\hspace*{4em}
\begin{subtable}{0.3\linewidth}
\centering
\begin{tabular}{|c|c|}
\hline
$o_1$ & $o_2$ \\ \hline
$o_2$ & $o_1$ \\ \hline
\end{tabular}
\caption*{$g_7$}
\label{tab:g7}
\end{subtable}
\quad
\hspace*{-4em}
\begin{subtable}{0.3\linewidth}
\centering
\begin{tabular}{|c|c|c|}
\hline
$o_1$ & $o_1$ & $o_2$\\ \hline
$o_3$ & $o_4$ & $o_3$\\ \hline

\end{tabular}
\caption*{$g_8$}
\label{tab:g8}
\end{subtable}
\quad
\hspace*{-3.5em}
\begin{subtable}{0.3\linewidth}
\centering
\begin{tabular}{|c|c|c|}
\hline
$o_1$ & $o_1$ & $o_2$\\ \hline
$o_4$ & $o_5$ & $o_2$\\ \hline
$o_4$ & $o_3$ & $o_3$ \\ \hline
\end{tabular}
\caption*{$g_9$}
\label{tab:g9}
\end{subtable}

\caption{Nine game forms. 
Alice and Bob choose rows and columns, respectively.\\
Forms $g_1$ - $g_6$ are tight, forms $g_7$ - $g_9$ are not; 
see Section \ref{tight}) for the definitions.}
\label{f1}
\end{figure*}

\subsection{Basic strategies and simple situations} 
%  A pair of strategies $(x,y)$  is called a {\em situation}.
Sets  $g(x) = \{g(x,y) \mid y \in Y\}$ and  $g(y) = \{g(x,y) \mid x \in X\}$
are called the {\em  supports} of strategies  $x \in X$  and   $y \in Y$, respectively.

A strategy is called {\em basic} or {\em minimal} if its support is not a proper subset of the support of any other strategy.
For example, in  $g_6$  the first strategies of Alice and Bob are basic, while the second are not;
in the remaining eight game forms all strategies are basic.
Moreover, any two strategies of a player, Alice or  Bob, have distinct supports.

A situation $(x, y)$ is called {\em simple}  if
$g(x)  \cap  g(y) =  \{g(x,y)\}$.
For example, all situations of  game forms  $g_1, g_2, g_8, g_9$  are  simple;  
% (such game forms are called  {\em rectangular});
in contrast, no situation is simple in  $g_7$;
in  $g_3$  all are simple, except three on the main diagonal;
in  $g_4$  all are simple, except the central one;
in  $g_6$  all are simple, except one with the outcome  $o_2$.

\medskip 

Game form  $g$  is called {\em rectangular} 
if all its situations are simple, or  
other words, each outcome  $o \in O$  
% of a rectangular game form  $g$  
fills a box, that is, 
$g^{-1}(o) = X' \times Y' \subseteq X \times Y$  
for some subsets  $X' \subseteq X$  and  $Y'  \subseteq Y$.
In Figure 1, game forms  $g_1,g_2,g_8,g_9$  are rectangular. 

\begin{remark} 
This concept can be generalized to 
the $n$-person case. 
In \cite{Gur82}, it was shown that 
an  $n$-person game form is the normal form of 
a positional structure with perfect information 
modeled by a tree if and only if this game form   
is tight and rectangular; see also 
\cite[Remark 2]{Gur75}, \cite{Gur84}, and \cite{Gur09} for more details. 
\end{remark} 

\subsection{Tight game forms} 
\label{tight}
Mappings  $\phi : X \rightarrow  Y$  and  $\psi : Y \rightarrow X$
are called {\em response strategies} of Bob and Alice, respectively.
The motivation for this name is clear:
a player chooses his/her strategy
as a function of a known strategy of the opponent.
As usual, $gr(\phi)$ and $gr(\psi)$ denote the graphs
of mappings  $\phi$  and  $\psi$  in  $X \times Y$.
Game form  $g$  is called {\em tight} if

\medskip

(j)  $g(gr(\phi)) \cap g(gr(\psi)) \neq \emptyset$
for every pair of mappings  $\phi$  and  $\psi$.

\medskip

It is not difficult to verify that in Figure 1
the first six game forms ($g_1 - g_6$) are tight, 
while the last three ($g_7 - g_9$) are not.

\medskip

In \cite{EF70,Gur73,Gur75,Gur89,Gur18,GN22}
the reader can find several equivalent properties characterizing tightness.
Here we mention some of them.

\medskip

(jjA) For every  $\phi : X \rightarrow  Y$
there exists a $y \in Y$  such that  $g(y) \subseteq g(gr(\phi))$.

\medskip

(jjB) For every $\psi : Y \rightarrow  X$
there exists a  $x \in X$  such that  $g(x) \subseteq g(gr(\phi))$.

\medskip

We leave to the careful reader to show that
(j) is equivalent to (jjA) and to (jjB) as well.

Properties (jjA) and (jjB) show that 
playing a game  $(g;u,w)$
with a tight game form  $g$, the players,
Bob and Alice, 
do not need non-trivial response strategies but
can restrict themselves by the constant ones, 
that is, by  $Y$ and $X$, respectively, 
at least in case of the zero-sum games.

\medskip

Given a  game form  $g : X \times Y \rightarrow O$,
introduce on the ground set $O$  two multi-hypergraphs  $A = A(g)$  and   $B = B(g)$
whose edges are the supports of strategies of Alice and Bob: % respectively:
$$A(g) = \{g(x) \mid x \in X\} \text{ and }  B(g) = \{g(y) \mid y \in Y\}.$$

By construction, the edges of $A$  and  $B$  pairwise intersect, that is,
$g(x) \cap g(y) \neq \emptyset$  for all  $x \in X$  and  $y \in Y$.
Furthermore,  $g$  is tight if and only if

\smallskip

(jjj) hypergraphs  $A(g)$  and  $B(g)$  are {\em dual}, that is, satisfy also the following two properties:

\begin{itemize}
\item [(jjjA)] for every  $O_A \subseteq  O$  such that
$O_A \cap  g(y) \neq \emptyset$  for all  $y \in Y$
there exists an  $x \in X$  such that  $g(x) \subset O_A$;
\item [(jjjB)] for every  $O_B \subseteq  O$  such that
$O_B \cap  g(x) \neq \emptyset$  for all  $x \in X$
there exists an  $y \in Y$  such that  $g(y) \subset O_B$.
\end{itemize}

\subsection{Tightness and solvability} 
\label{tight-solv} 
Consider the following properties of a game form: 

\medskip 

(i) NS, (ii) zero-sum-solvability, (iii) win-lose-solvability, and  (iv) tightness. 

\medskip 

Implications (i) $\; \Rightarrow \;$  (ii)  $\; \Rightarrow \;$ (iii)
are immediate from the definitions. 

Also  (iii)  $\; \Rightarrow \;$ (iv) is easily seen. 
Indeed, if  $g$  is not tight then there exist response strategies 
$\phi$  and  $\psi$  such that 
$g(gr(\phi)) \cap g(gr(\psi)) = \emptyset$.
Define the zero-sum reward  $r$  such that 
$r(o) = -1$   for  $o \in g(gr(\phi)$, 
$\; r(o) = 1$  for  $o \in g(gr(\phi)$, and 
$r(o) = \pm 1$, arbitrarily,  for the remaining outcomes, if any. 
%  $o \in O \setminus [g(gr(\phi)) \cup g(gr(\psi))]$, if any. 
Set  $r^A = r$  and  $r^B = -r$.
Obviously, $-1 = min max < max min= 1$  in the obtained 
win-lose game. Hence, it has no saddle point.

\medskip 

Implication (iv) $\; \Rightarrow \;$ (ii) is implicit 
in \cite{EF70} and explicit in \cite{Gur73}. 
Finally, (iv) $\; \Rightarrow \;$ (i)  
appears in \cite{Gur75}; see also \cite{Gur89}. 
Thus, all four properties  (i-iv) are equivalent.

\medskip 

However, for the three-person case tightness 
is no longer sufficient \cite{Gur75} 
nor necessary \cite{Gur75,Gur89}  for NS. 

\medskip 

Recently, implication  (iv) $\; \Rightarrow \;$ (i)  
was strengthened and its proof simplified in \cite{GN22,GN22A}.  

Given a tight game game form  $g : X \times Y \rightarrow O$ 
and reward functions  
$r^A : O \rightarrow \RR$  and   $r^B : O \rightarrow \RR$, 
game  $(g, r^A, r^B)$  has a simple NE in basic strategies, 
that is,  situation  $(x,y)$ is simple and strategies  $x,y$  are basic. 
Moreover, there exist some special NE $(x^0, y^*)$
defined by a lexicographically safe strategy $x^0$  of Alice, 
which depends only on her cost function  $r^A$, while 
Bob's cost function  $r^B$  is irrelevant. 
This concept is a refinement of the classic safe (max min) strategy.
Then, Bob's NE strategy $x^*$ maximizes  $r^B$  
over set  $g(x^0)$, which is the support of  $x^0$. 
Of course, Alice and Bob can be swapped. 
Thus, we obtain two sets of NE:  NE-A  and  NE-B.  
These two sets coincide in the zero-sum case 
and in some other cases too; 
for example, when game $(g, r^A, r^B)$  has a unique NE. 
However, in general, NE-A and  NE-B differ. 
Furthermore, a pair of 
lexicographically safe strategies   
$(x^0, y^0)$  is not necessarily a NE. 
See more details in  \cite{GN22,GN22A}.

\medskip 

These statements are constructive: 
a polynomial algorithm determining NE-A and NE-B is suggested. 
This is trivial when a tight game form  $g$  is explicit, but 
the algorithm works in a more general case, when  $g$  is given 
by a polynomial oracle $\cO$ such that 
the size of  $g$  is exponential in the size of $\cO$. 
In the next subsection we consider an example of such oracle.

\subsection{Deterministic graphical multi-stage game structures}
\label{DGMSGS}
Let $\Gamma = (V,E)$ be a directed graph (digraph) whose vertices
and arcs are interpreted as positions and moves, respectively.
Furthermore, denote by  $V_T$  the set of terminal positions, of out-degree zero,
and by  $V_A, V_B$  the  positions of positive out-degree,
controlled by Alice and Bob, respectively.
We assume that  $V = V_A  \cup  V_B \cup V_T$  is a partition.
A strategy $x \in X$  of Alice
(resp., $y \in Y$  of Bob) is a mapping that
assigns to each position  $v \in V_A$
(resp., $v \in V_B$) an arbitrary move from this position.
An initial position  $v_0 \in V_A \cup V_B $ is fixed.
Each situation $(x,y)$  defines a unique a walk
that begins and  $v_0$  and then follows the decisions made by  $x$  and $y$.
This walk $P(x,y)$ is called a {\em play}.
Each play either terminates in  $V_T$  or is infinite.
In the latter case, it forms a ``lasso":
first, an initial path, which may be empty, and 
then a directed cycle (dicycle) repeated infinitely
(This holds, because we restrict players
by their stationary strategies,  that is,
a move may depend only on the current position but
not on previous positions and/or moves).

The positional structure defined above can also be represented in normal form.
We introduce a game form  $g : X \times Y \rightarrow O$,
where, as before, $O$  denotes a set of outcomes.
Yet, there are several ways to define this set.
One is to ``merge" all infinite
plays (lassos) and consider them as a single outcome  $c$,
thus, setting  $O = V_T \cup \{c\}$.
This model was introduced by Washburn \cite{Was90}
and called {\em deterministic graphical game structure} (DGGS).

The following generalization was suggested in \cite{Gur18}.
Digraph  $\Gamma$  is called {\em strongly connected}  if
for any  $v, v' \in V$  there is a directed path from  $v$  to  $v'$
(and, hence, from  $v'$ to $v$, as well).
By this definition, the union of two strongly connected digraphs
is strongly connected  whenever they have a common vertex.
A vertex-inclusion-maximal strongly connected induced subgraph of $\Gamma$
is called its {\em strongly connected component} (SCC).
In particular, each terminal position $v \in V_T$ is an SCC.
It is both obvious and well-known that
any digraph  $\Gamma = (V, E)$  admits a unique decomposition into SCCs:
$\Gamma^o = \Gamma[V^o] = (V^o, E^o)$  for $o \in O$, where
$O$  is a set of indices.
Furthermore, partition  $V = \cup_{o \in O} V^o$  can be constructed in time linear
in the  size of  $\Gamma$, that is, in $(|V| + |E|)$.
It has numerous applications; see \cite{Sha81,Tar72} for more details.
One more was suggested in \cite{Gur18}.
For each  $o \in O$, contract the SCC  $\Gamma^o$  into a single vertex  $v^o$.
Then, all edges of  $E^o$ (including loops) disappear and
we obtain an acyclic digraph  $\Gamma^* = (O, E^*)$.
The set  $O$  can be treated as the set of outcomes.
Each situation  $(x,y)$ uniquely defines a play $P = P(x,y)$.
This play either comes to a terminal  $v \in V_T$  or forms a lasso.
The cycle of this lasso is contained in an SCC  $o$  of $\Gamma$.
Each terminal is an SCC as well.
In both cases an SCC  $o \in O$ is assigned to the play  $P(x,y)$.
Thus, we obtain a game form  $g : X \times Y \rightarrow O$,
which is the normal form of the
{\em multi-stage DGGS} (MSDGGS) defined by  $\Gamma$.

An SCC is called {\em transient} if it is not a terminal
and contains no dicycles.
No play can result in such SCC; or in other words
it does not generate an outcome.
For example, $O = V_T$  in an acyclic digraph, while all remaining SCC are transient.

DGGS and MSDGGS can be viewed as polynomial oracles 
the corresponding game forms  $g'$  and  $g$, respectively. 
Note that the size of game forms may be exponential in the size of these oracles. 
Note also that  $g'$  is  obtained from  $g$  by merging some outcomes.
Namely, all outcomes corresponding to the non-terminal SCCs
are replaced by a single outcome $c$.
Obviouslly, merging outcomes respects tightness.
Thus, it is enough to verify it for MSDGGSs,
As we know, it is sufficient to prove the win-lose solvability.
For DGGS it was done in \cite{Was90}; see also
\cite[Section 3]{BG03}, \cite{AHMS08}, \cite[Section 12]{BGMS07}.
The result was extended to MSDGGS in \cite{Gur18}.
All proofs were constructive,
the corresponding win-lose games were solved in time
polynomial in the size of  $\Gamma$.

\medskip

For reader's convenience, we briefly sketch here the proof of from \cite{Gur18}.
Consider a win-lose game $(g;r^A,r^B)$ with game form  $g = g(\cO)$ 
generated by a MSDGGS oracle $\cO$.
Let  $O = O_A \cup  O_B$  denote 
the partition of  $O$  into two sets of outcomes: 
winning for Alice and Bob, respectively. 
We would  like to apply the Backward Induction,
yet, digraph $\Gamma$   may have dicycles.
So we modify Backward Induction to make it work in presence of dicycles.
Recall that $O$  is the set of SCCs of  $\Gamma$ and  $\Gamma^* =
(O, E^*)$  is acyclic.
Consider an SCC  $o = \Gamma' = (V',E')$  in  $\Gamma$
that is not terminal, but each move  $(v',v)$
from a position  $v' \in V'$
either ends in a terminal $v \in V_T$, or
stays in  $\Gamma'$, that is, $v' \in V'$.
Wlog assume that $o \in O_A$, that is,
Alice wins if the play cycles in $\Gamma'$.
Then Bob wins in a position $v' \in V'$ if and only if
he can force the play to come to a  terminal $v \in O_B$  and
Alice wins in all other positions of $V'$.
Note that for Alice it is not necessary to force the play to come to $O_A$,
it is enough if it cycles in $\Gamma'$.
Thus, every position of  $\Gamma'$ can be added either to  $O_A$  or to  $O_B$.
Then we eliminate all edges  $E'$  of  $\Gamma'$ and repeat
until the initial position  $v_0$  of $\Gamma$  is evaluated.
This procedure proves solvability of game form  $g(\cO)$ and
solves a win-lose game $(g;O_A,O_B)$  in time linear in the  size of $\cO = \Gamma$.

\medskip 

More polynomial oracles for (finite two-person) tight game forms 
can be found in \cite{GK18,GN22}. 

\section{Zero-sum-solvability of tight $v$-forms} 
Here we survey results of \cite{Gur90}.  
\subsection{Main concepts and theorem} 
A {\em vector game form of type} $v$  
(or $v$-{\em form)} is a mapping 
$g_v : X \times Y \rightarrow W$, 
where  $W \subset \RR^m$  
is a finite set of real $m$-vectors  
$\{w = (w_1, \ldots, w_m)  \mid  w \in W\}$.

Given an arbitrary real utility $m$-vector $u = (u_1, \ldots, u_m)$, 
we define a real-valued reward function $r : W \rightarrow \RR$, where 
$r(w) = (u, w) = u_1 w_1 + \ldots u_m w_m$  for each  $w \in W$. 

In the zero-sum case we assume that  
$r^A = r$ and  $r^B = -r$; 
Alice is the maximizer and  Bob is the minimizer.
In general case, we introduce 
$u^A : W \rightarrow \RR$ and  $u^B : W \rightarrow \RR$  separately, 
set  $r^A(w) = (u^A, w)$  and  $r^B(w) = (u^B, w)$  
for all vectors  $w \in W$, and assume that  
both players are maximizers. 

\medskip 

% \subsection{Tightness of $v$-forms}
A $v$-form  $g_v$  is called {\em tight} if 
$$Conv Hull \{g_v(x, \phi(x)) \mid x \in X\} \;  \cap  \; 
 Conv Hull \{g_v(\psi(y), y) \mid y \in Y\} \neq \emptyset
\;\;\; \forall \;  \phi : X \rightarrow Y, \; \psi : Y \rightarrow X;$$

\noindent 
in (other) words, for an arbitrary pair of response strategies  
$\phi : X \rightarrow Y$  of Alice  and  $\psi : Y \rightarrow X$  of Bob,   
convex hulls of the corresponding two sets of vectors from their graphs, 
% \newline 
$\; \{g_v(\phi) = \{g_v(x, \phi(x)) \mid x \in X\}, \; \{g_v(\psi) = \{g_v(\psi(y), y) \mid y \in Y\} \subseteq W \;$,  intersect in  $\RR^m$. 
 
\medskip 

A game form $g$  can be viewed as a $v$-form  $g_v$  
with only unit vectors 
(one entry $1$ and all others $0$).
For any two sets of such vectors, 
their convex hulls are disjoint in $\RR^m$  
if and only if the sets are disjoint.  
Thus, definitions of tightness 
for game forms and for vector game forms agree, 
that is,  $g$  and $g_v$  are tight simultaneously. 

\smallskip 

The next theorem extends the criterion of 
zero-sum-solvability of Section 1. 

\begin{theorem} 
\label{t1} 
(\cite{Gur90}) 
Tightness and zero-sum-solvability of $v$-forms are equivalent. 
\end{theorem}

{\bf Proof.}
Suppose that  $g_v$  is not tight. 
Then there exist  $\phi$  and  $\psi$  such that 
\newline
$g_v(\phi) \cap  g_v(\psi) = \emptyset$. 
Two disjoint convex sets in $\RR^m$  can be separated by a hyperplane; 
\newline
in other words, there exists a vector $u \in \RR^m$  such that  
$(u, g_v(x, \phi(x)) < (u, g_v(\phi(y), y)$  
for every  $x \in X$  and $y \in Y$.  
Then, $max min < min max$  in the zero-sum game $(g, u)$ and, 
hence, it has no saddle point. 
Thus, $g_v$  is not zero-sum-solvable. 

\medskip 

Conversely, suppose that  $g_v$  is not zero-sum-solvable,  
that is, for some  $u \in \RR^m$, 
the zero-sum game  $(g,u)$ has no saddle point and. 
Then,  $max min < min max$  in this game. 
Consider arbitrary best response strategies 
$\phi: X \rightarrow Y$  and  $\psi: Y \rightarrow X$ 
of Bob and Alice, respectively. 
By definition, these two strategies 
guarantee $max min$ and  $min max$, respectively. 
Hence, $g_v(\phi) \cap  g_v(\psi) = \emptyset$. 
Thus, $g_v$  is not tight.
\qed 

\begin{remark} 
A tight game form  $g : X \times Y \rightarrow O$  
is injective if and only if  $|X| = 1$  or   $|Y| = 1$. 
In contrast, a tight $v$-form  $g_v$  may be injective 
for any sizes of  $X$  and  $Y$.

\medskip 

As we know, NS and zero-sum-solvability are equivalent 
for game forms. In contrast, NS of a $v$-form 
does not follow from its zero-sum-solvability. 
For example,  mean payoff games are  zero-sum-solvabile 
but not NS \cite{Gur88}; see next subsection for more details.

\medskip

Verifying tightness of an explicitly given game form 
$g$ is an important open problem. 
\newline 
A quasi-polynomial algorithm was suggested  
by Fredman and Khachiyan \cite{FK96}; see also  \cite{GK99}. 
``Almost obviously" 
verifying tightness of an explicitly given 
$v$-form is NP-complete. 
Yet, a proof is required.
\end{remark} 

\subsection{Mean payoff games} 
Consider the model of Section  \ref{DGMSGS}. 
Add a loop to every terminal position  $v \in V_T$, if any. 
Then, the obtained graph  $G = (V,E)$  
has no terminal positions. 

Set  $O$  of outcomes consists of all dicycles 
(dicycles) of $G$.
To each dicycle  $C$  assign an  $m$-vector of weights 
$w(C) = (w_e \mid e \in E)$  
such that   $w_e = 1/k$  for  $e \in C$  and 
$w(e) = 0$  for  $e \not\in C$. 
Here  $m = |E|$ is the number of directed edges of $G$ and 
$k = k(C) = |C|$  is the length of dicycle $C$. 
Note  that sum of entries of any vector  
$w(C)$  equals  $k(1/k) = 1$. 
Note also that  $k$  may take any positive integer value: 
$k=1$  if $C$  is a loop and  $k = 2$  if  
$C$ is formed by a pair of oppositely directed edges.

Given a utility vector 
$u = (u(e) \mid e \in E) = (u_1, \ldots, u_m)$, 
a real-valued reward function $r : W \rightarrow \RR$ 
is the scalar product 
$r(w) = (u, w) = u_1 w_1 + \ldots u_m w_m$  
for each  $w \in W$. 

As before, in the zero-sum case we assume that  
$r^A = r, r^B = -r$; 
Alice is the maximizer and  Bob is the minimizer.
In general case we introduce 
$u^A : W \rightarrow \RR$  and  $u^B : W \rightarrow \RR$  separately,  
set  $r^A(w) = (u^A, w), \; r^B(w) = (u^B, w)$  
for all vectors  $w \in W$, and assume that  both players are maximizers. 

Since  $V_T = \emptyset$, 
each play $P = P(x,y)$ is infinite;  
it forms a ``lasso" consisting of 
an initial path, which may be empty, and
a dicycle  $C = C(x,y)$  repeated infinitely. 
Thus, the effective payoff of a player 
in situation $(x,y)$ is 
the average local payoff along $C(x,y)$, that is, 
$r(C) = (u, w) = |C|^{-1} \sum_{e \in C} u_e$, 
where  $u$  is $u^A$ or $u^B$ for Alice and Bob, respectively.
These definitions justify the name "mean payoff games". 

Zero-sum-solvability of these games was proven 
by Moulin \cite{Mou76}  for complete bipartite digraphs, 
by  Ehrenfeucht and Mycielski \cite{EM79} 
for any bipartite digraphs, and by 
Gurvich, Karzanov, and Khachiyan \cite{GKK88} for arbitrary digraphs. 

Thus, by Theorem \ref{t1}, the corresponding 
$v$-forms are tight. 
However, no direct proof of tightness is known and 
it is a challenge to obtain one. 

\medskip  

Furthermore, NS does not hold. 
An example of NE-free mean payoff game was constructed 
in \cite{Gur88} for the complete bipartite digraph $3 \times 3$. 
In a way, this example is minimal: 
NS holds for the complete bipartite digraphs 
$a \times b$  with  $a \leq 2$  or  $b \leq 2$  \cite{Gur90A}. 

Thus, unlike game forms, 
for $v$-forms NS and zero-sum-solvability are not equivalent. 

\subsection{Mean payoff games with positions of chance}
Replace partition  $V = V_A \cup V_B$  by 
$V = V_A \cup V_B \cup V_R$  
by allowing positions $V_R$ with random moves,  
with a given probabilistic distribution 
on the edges going from each  $v \in V_R$. 
Introduce also a probabilistic distribution  
over  $V$,  where  $p(v)$  is the probability 
that the game begins in  $v$. 
In particular, an initial position $v_0$  may be fixed, 
which means that  $p(v_0) = 1$  and  $p(v) = 0$ for 
all  $v \in V \setminus \{v_0\}$. 

Then, each situation  $(x,y)$  determines 
a Markov's chain  $M(x,y)$  
rather than a unique play  $P(x,y)$ 
in the obtained digraph  $G = (V,E)$. 
In accordance with Markov's theory, 
this chain has a limit probabilistic distribution 
$p : E \rightarrow \RR$, with  $p(e) = p_e \geq 0$  
for all  $e \in E$ 
and $\sum_{e \in E} p_e = 1.$  
Setting  $w_e(x,y) = p_e(x,y)$  
for all  $(x,y)$ and  $e \in E$  
we obtain a  $v$-form. 

\medskip 

This model was suggested in \cite{GKK88}, 
where BW-games and  BWR-games were introduced.  
(Bob and Alice's positions were called 
Black (B) and White (W), respectively.)   
In fact, BWR-games are computationally equivalent 
\cite{BEGM13,BEGM17} to classic 
stochastic games with perfect information 
and zero stop probability, which were introduced much earlier,
in 1957, by Gillette \cite{Gil57}, who proved their zero-sum-solvability. 
This proof is not simple; it is based on 
the famous Hardy and Littlewood Tauberian theorem; 
its conditions were not accurately verified in \cite{Gil57} 
yet; so the proof was completed only in 12 years  
by Liggett and Lipman \cite{LL69}.

\smallskip

Thus, by theorem \ref{t1}, $v$-forms are tight,  
yet, no direct proof of their tightness is known.

\begin{remark} 
In $v$-forms considered in this 
and in the previous subsections, 
all vectors $w \in W$  are non-negative, 
$w_e \geq 0$  for all  $e \in E.$ 
Yet, Theorem \ref{t1} holds for arbitrary real vectors.
%% w = ur,  r  may be negative. So what?  still  w \geq 0 
\end{remark} 

In addition to mean payoff games, 
some other zero-sum-solvabile cases 
are known in stochastic game theory. 
An infinite family of effective payoffs, 
called  $k$-total, was considered in \cite{BEGM17A} 
for any integer  $k \geq 0$. 
For example, $k=0$  and  $k=1$ are associated 
with mean and {\em total} \cite{BEGM18} effective payoffs, respectively.
For each  $k$,  the corresponding $v$-forms are NS \cite{BEGM17A} 
and, hence, tight, by Theorem \ref{t1}. 
Yet, no direct proof of tightness is known. 

\section{Nash-solvability of tight $v^+$-forms}

\subsection{Main concepts} 
{\bf Vector game forms $g^+_v$.}  
We modify the concept of $v$-forms 
by requiring some extra properties.
A {\em vector game forms of type} $v^+$  
(or $v^+$-{\em form)} is a mapping 
$g^+_v : X \times Y \rightarrow \{W \cup \{w^c\}\}$, 
where  $W \subset \RR^m$  
is a finite set of real 
{\em non-negative} and {\em non-zero} $m$-vectors,   
$\{w = (w_1, \ldots, w_m)  \mid  w \in W\}$  with  
$w_i \geq 0$  for all $i = 1, \ldots, m$ and 
$w_i > 0$  for at least one $i \in \{1, \ldots, m\}$;   
while  $w^c$  is as a special $m$-vector  
all $m$ coordinates of which are  $+\infty$.

\bigskip 

{\bf Weak rectangularity.}
We will also assume that mapping $g^+_v$ is 
{\em weakly rectangular}, that is,
$(g^+_v)^{-1} (w) = X' \times Y'$, 
where $X' \subseteq X$  and  $Y' \subseteq Y$, 
or in other words, 
each vector $w \in W$  fills a box in $X \times Y$. 
In contrast, $w^c$  
fills the rest of  $X \times Y$ 
and may be not a box. 

\bigskip 

{\bf Local and effective costs.} 
Let 
$u^A = (u^A_1, \ldots, u^A_m)$  and 
$u^B = (u^B_1, \ldots, u^B_m)$  
be strictly positive cost vectors 
of Alice and Bob, respectively;  
$u_i^A > 0$  and  $u_i^B > 0$ 
for all  $i = 1, \ldots m$.

Now we assume that both players are minimizers, 
rather than maximizers. 
Respectively, we replace rewards by costs.
Define real-valued cost functions 
$r^A : W \rightarrow \RR$   and 
$r^B : W \rightarrow \RR$  
as scalar products 
$r^A(w) = (u^A, w)$  and  
$r^B(w) = (u^B, w)$, respectively, 
for each  $w \in W$. 

Note that both functions are strictly positive. 
Furthermore, set 
$r^A(w^c) = r^B(w^c) = +\infty$. 

Then, effective costs 
$r^A : X \times Y \rightarrow \RR$ and 
$r^B : X \times Y \rightarrow \RR$
for Alice and Bob 
are defined as  
$r^A(x,y) = r^A(g^+_v(x,y))$  and 
$r^B(x,y) = r^B(g^+_v(x,y))$  
for each situation  $(x,y)$, 
including the case when  $w(x,y) = w^c$.

\bigskip 

{\bf Degenerate and non-degenerate situations, NE, strategies, and game forms.}

Given a $v^+$-form $g^+_v$, 
\begin{itemize} 
\item {} situation $(x,y)$ is called {\em degenerate}  
if  $g^+_v(x,y) = w^c$;
\item {} strategy $x$ of Alice  (respectively, $y$ of Bob) is called {\em degenerate} 
if  $g^+_v(x, y) = w^c$  for each $y \in Y$ 
(respectively, for each $x \in X$);
% \item {} strategy $y$ is called {\em degenerate}  
% if  $g^+_v(x, y) = w^c$  for each $x \in X$;
\item {} $v^+$-form $g^+_v$ itself 
is called {\em degenerate}  
if  $g^+_v(x, y) = w^c$  for each $x \in X$ and $y \in Y$.
\end{itemize} 

The following statements are obvious: 

Degenerate situation  
$(x,y)$ is a NE  if and only if 
both strategies  $x$  and $y$  are degenerate. 

Each situation of a degenerate $v^+$-form is a degenerate NE. 
This case is trivial.

\bigskip 

{\bf Possibly best response (PBR) strategies.} 
A response strategy  $\phi : X \rightarrow Y$ of Bob 
is called a {\em PBR}   
if there exists a cost vector  $u^B$  such that 
$y = \phi(x)$  is a best response to  $x$  for each 
Alice's strategy $x$. 

Without any loss of generality 
(wlog), we can additionally require uniqueness 
of the best response vector  $w(x) = g^+_v(x, y(x))$  
in the support  $g(x) = \{g(x,y) \mid y \in Y\}$ 
for all  $x \in X$.
Indeed, if uniqueness does not hold, 
it can be achieved by a perturbation of  $u^B$, 
sufficiently small 
to respect the best response strategy  $y = \phi(x)$. 

\smallskip

Similarly, we define PBRs $\psi : Y \rightarrow X$ 
of Alice. 

\bigskip 

{\bf Tightness} 
A $v^+$-form $g^+_v$ is called {\em tight} 
if supports of any two PBRs  $\phi$ and $\psi$ 
of Alice and Bob intersect:  

$$\{g^+_v(x, \phi(x)) \mid x \in X\} \cap  
  \{g^+_v(\psi(y), y) \mid y \in Y\} \neq \emptyset.$$
  
\subsection{Main theorem} 
Tightness and NS are equivalent for  $v^+$-forms.

\begin{theorem} 
\label{t2} 
A $v^+$-form  $g^+_v$  is tight if and only if it is NS. 

Moreover, if  $g^+_v$  is tight and 
Alice {\bf or} Bob has no degenerate strategy  
then for any cost vectors  $u^A$  and  $u^B$ 
the corresponding game  $(g^+_v, u^A, u^B)$  
has a non-degenerate NE. 
\end{theorem} 

{\bf Proof} 
As we already mentioned, 
a degenerate $v^+$-form is tight; furthermore, 
every its situation is a degenerate NE. 
This case is trivial. 

\medskip

Obviously, Alice (respectively, Bob) 
has a degenerate strategy if and only if  

\smallskip 

$w^c \in \{g^+_v(x, \phi(x)) \mid x \in X\}$ 
(respectively, 
$w^c \in \{g^+_v(\psi(y), y) \mid y \in Y\}$.  

\smallskip 
\noindent
Here $\phi$ and $\psi$  
are some response strategies of Bob and Alice, respectively. 

\medskip{}
Suppose that both players   
have degenerate strategies,  
Alice  $x$  and Bob  $y$.
Then, for any  $u^A$ and  $u^B$  we have: 

\begin{itemize} 
\item[] Both sets 
$\{g^+_v(x, \phi(x)) \mid x \in X\}$  and   
$\{g^+_v(\psi(y), y) \mid y \in Y\}$ 
contain  $w^c$ (in fact, only  $w^c$).  
Hence, their intersection is not empty. 
Thus, $g^+_v$  is tight. 
\item[] Game $(g^+_v, u^A, u^B)$  has a degenerate NE 
for any cost vectors $u^A$ and $u^B$.  
\end{itemize}
The statement holds in this case too. 

\medskip 

For the rest of the proof, assume that Alice and Bob 
have no degenerate strategies. 
\newline 
Then 
$w^c \not\in 
\{g^+_v(x, \phi(x)) \mid x \in X\} \cup  
\{g^+_v(\psi(y), y) \mid y \in Y\}$ 
% \smallskip 
% \noindent 
for any cost vectors  $u^A$  and  $u^B$ 
and corresponding best response strategies 
$\phi$ and $\psi$  of Bob and Alice, respectively.  

\medskip

Assume that  $g^+_v$  is tight. Then 
$\{g^+_v(x, \phi(x)) \mid x \in X\} \cap  
\{g^+_v(\psi(y), y) \mid y \in Y\} \neq \emptyset$ 
and, hence, there exists a vector 
$w^* \in W \cap 
\{g^+_v(x, \phi(x)) \mid x \in X\} \cap  
\{g^+_v(\psi(y), y) \mid y \in Y\}$.

\smallskip  

Recall that vector game form $g^+_v$ is weakly rectangular. 
Hence, for any  $u^A$  and  $u^B$, 
there exist strategies  $x$  and $y$  such that 
$y = \phi(x), \; x = \psi(y)$, and  $g^+_v(x,y) = w^*$, 
where  $\phi$ and $\psi$ are best response strategies 
of Bob and Alice, respectively. 
In other words, situation  
$(x, y)$  is a non-degenerate NE 
in game  $(g^+_v(x,y), u^A, u^B)$. 

\medskip 

Conversely, assume that $g^+_v$  is not tight. 
Then there exist PBR  $\phi$ and  $\psi$  such that 
$\{g^+_v(x, \phi(x)) \mid x \in X\} \cap  
\{g^+_v(\psi(y), y) \mid y \in Y\} = \emptyset$. 
Obviously, for the corresponding  
\newline 
$u^A$  and $u^B$ 
game    $(g^+_v(x,y), u^A, u^B)$  has no NE. 
Recall that  $(x,y)$  is a NE if and only if 
$y$  is a best response to  $x$  and 
$x$  is a best response to  $y$.
\qed 

\medskip

If a player has no degenerate strategies  
then delete them from the opponent's set 
of strategies as well. 
Obviously, this operation respects tightness 
and, hence, NS too. 

\subsection{Asumability conditions}
Verifying tightness of a $v^+$-form looks very difficult, 
although no accurate results abut its complexity is known. 
In particular, it is not clear how to check that 
mapping  $\phi : X \rightarrow Y$  is a PBR. 
However, the following, pretty strong, conditions are necessary. 
Consider a subset  $X^* \subseteq X$  and two mappings 
$\phi' : X^* \rightarrow Y$  and  $\phi'' : X^* \rightarrow Y$.
Assume that all vectors of $W$  are pairwise distinct 
and also that  $2 |X^*|$ vectors  
% \newline  
$\{g^+_v(x, \phi'(x)), g^+_v(x, \phi''(x)) \mid x \in X^*\} 
\subseteq  W$  
are pairwise distinct, for all  $x \in X^*$. 

\begin{proposition} 
If  $|X^*| \geq 1$  and  
$\;\; \sum_{x \in X^*} g^+_v(x, \phi'(x)) >    
 \sum_{x \in X^*} g^+_v(x, \phi''(x))$ 
 
\smallskip 
\noindent 
then mapping  
$\; \phi' : X^* \rightarrow Y$  cannot be extended 
to a PBR  $\phi : X \rightarrow Y$. 

\medskip 

If  $|X^*| \geq 2$  and  
$\;\; \sum_{x \in X^*} g^+_v(x, \phi'(x)) =     
 \sum_{x \in X^*} g^+_v(x, \phi"(x))$ 
 
\smallskip 
\noindent 
then neither  
$\; \phi' : X^* \rightarrow Y$  nor 
$\; \phi'' : X^* \rightarrow Y$ 
can be extended 
to a PBR  $\phi : X \rightarrow Y$. 
\end{proposition} 

{\bf Proof.} 
The first statement is obvious, since 
Bob's cost is  $r^B(w) = (u^B, w)$  
all entries are non-negative, and Bob is the minimizer. 

For the second statement we should recall 
that  wlog we cab assume uniqueness of the best response vector  
$w(x) = g^+_v(x, y(x))$  
in the support  $g(x) = \{g(x,y) \mid y \in Y\}$ 
for all  $x \in X$; see above. 

Furthermore, without any loss of generality 
(wlog), we can additionally require uniqueness 
of the best response vector  $w(x) = g^+_v(x, y(x))$  
in the support  $g(x) = \{g(x,y) \mid y \in Y\}$ 
for all  $x \in X$. 
Indeed, if uniqueness does not hold, 
it can be achieved by a perturbation of  $u^B$, 
sufficiently small 
to respect the best response strategy  $y = \phi(x)$. 
\qed 

\medskip

Of course, similar necessary conditions hold 
for Alice's PBRs as well. One should just replace  
$x, X$, and $\phi$  by  $y, Y$, and  $\psi$, respectively. 

\medskip

It is open whether the above asumability conditions 
only necessary or necessary and sufficient 
for a response strategy to be a PBR. 

\section{Shortest path games and bi-shortest path conjecture}
% Outcomes are paths. No tightness 
% \subsection{Bi-shortest path conjecture.} 
We formulate a conjecture from graph theory 
\cite{Gur21,BFGV22}  that is equivalent 
to NS of the finite two-person vector $v^+$-forms,  
which correspond to the so-called 
shortest path games with positive local costs.  
For the three-person case this conjecture fails \cite{GO14}. 

\subsection{Definitions and statement of the conjecture}
Let  $G = (V,E)$   be a finite digraph 
with two distinct vertices  $s, t \in V$.   
We assume that 
\begin{itemize} 
\item[(j)] every vertex  $v \in V \setminus \{t\}$  has an outgoing edge, 
while  $t$  has not;  
\item[(jj)] $G$ contains a directed path from  $s$ to $t$;  
\item[(jjj)] every edge  $e \in E$  belongs to such a path. 
\end{itemize} 
If (j) fails for  $v$ we merge  $v$ and  $t$; 
if (jjj)  fails for  $e$  we delete  $e$  from  $E$.

\medskip   

Given a partition  $V \setminus \{t\} = V_A \cup V_B$  
with non-empty  $V_A$  and  $V_B$, 
assign an ordered pair of positive real numbers  
$(u^A(e), u^B(e))$  to every $e \in E$.

Fix a mapping  $x$  that assigns 
to each  $v \in  V_A$  an edge  $e \in E$  going from  $v$.
Delete all other edges going from  $v$. 
%  thus getting a  graph  $G(s_i)$,   
In the obtained digraph find a directed shortest path 
(SP) from  $s$  to  $t$,  assuming that  
$u^B$  are the lengths (or costs) of the edges  $e \in E$. 
One can use, for example, Dijkstra's SP algorithm. 
Swapping  $A$  and  $B$  we obtain two sets of 
directed  $(s,t)$-paths. 

We  conjecture that these two sets intersect, 
that is, have an $(s,t)$-path in common, and 
call this {\em bi-shortest path (Bi-SP) conjecture}. 

\medskip 

Wlog, we can assume that all  $(s,t)$-paths 
have pairwise different  lengths, 
which can be achieved 
by small perturbations of  $u^A$  and $u^B$. 
Then, a shortest path is unique. 

\medskip  

It may happen that some mappings  $x$ or $y$ 
leave no $(s,t)$-path.  
Then, we choose nothing.

Let us slightly modify the procedure  
choosing in this case some symbolic path $c$. 
Then we obtain a weak version of the Bi-SP conjecture.  
Indeed, if  two sets of $(s,t)$-paths have only $c$ in common    
then the Bi-SP conjecture fails, but the weak Bi-SP one holds. 
   
\medskip

Wlog, we can restrict ourselves by bipartite graphs 
with parts $(V_A, V_B)$. 
Indeed, if  $E$  contains an edge  $e = (u,w)$   
such that both  $u$ and  $w$  are in $V_A$  
(respectively, in  $V_B$)  
we subdivide  $e$  by a vertex  $v \in V_B $  
(respectively, by  $v \in V_A$)   
into two edges  $e' = (u,v)$  and  $e'' = (v,w)$  
choosing some lengths  
$u^D(e') > 0$  and  $u^D(e'') > 0$  such that 
$u^D(e) = u^D(e') + u^D(e'')$, where  $D = A$  or  $D = B$.

\subsection{Finite $n$-person shortest path games} 
{\bf Players, positions, moves, and local costs.}  
Given a finite digraph  $G =(V, E)$  
satisfying above assumption (j, jj, jjj), 
let us generalize case of $n=2$ players 
and consider an arbitrary integer  $n \geq 2$.
Partition vertices into  $n$  non-empty subsets 
$\{V \setminus t\} = V_1 \cup \ldots \cup V_n$  and 
assign a positive real number 
$u^i(e)$  to every player  $i \in I$  and edge  $e \in E$. 
Consider the following interpretation: 
$I = \{1, \ldots, n\}$  is a set of {\em players}, 
$V_i$ is the set of {\em positions} controlled by player $i \in I$; 
furthermore, $s = v_0$  and  $t = v_t$  are respectively 
the {\em initial} and {\em terminal} positions; 
$e \in E$  is a {\em (legal) move},  
$u^i(e)$  is the cost of move $e  \in E$  for player  $i \in I$   
called the {\em local  cost}. 
% , and  $(u^i = u^i(e) \mid e \in E = (u^i_1, \ldots, u^i_m)$  

\medskip 

{\bf Strategies, plays, and effective costs.}
A mapping  $z_i$  that  assigns a move  $(v, v')$  
to each position  $v \in V_i$  is a strategy of player  $i \in I$. 
\footnote{We restrict ourselves and all players  
to their pure stationary strategies;   
no mixed or history dependent ones are considered in this paper.}  
Each {\em strategy profile} (also called a {\em situation})   
$z = (z_1, \ldots, z_n) \in Z_1 \times \ldots \times Z_n = Z$   
uniquely defines a play  $p(z)$, that is, a walk in $G$  
that begins in the initial position  $s = v_0$  and 
goes in accordance with the choice of  $z$  
in every position that appears. 
Obviously, $p(z)$  either terminates in  $t = v_t$  or cycles; 
respectively, it is called a terminal or a cyclic play. 
Indeed, after play  $p(z)$  revisits a  position, 
this play will repeat its previous moves, thus, making a ``lasso", 
because all strategies in situation  $z$  are stationary. 

\smallskip

The effective cost of play $p(z)$  
for player  $i \in I$ is additive, that is,  
$$r^i(p(z)) = \sum_{e \in p(z)} u^i(e) 
\;\;\;  \text{if $p(z)$  is a terminal play;}$$    
$$r^i(p(z))= +\infty  \;\;\; \text{if  $p(z)$  is a cyclic play.}$$
In other word,  each player $i \in I$  pays the local cost  $u^i(e)$  
for every move $e \in p(z)$. 
Since a cyclic play  $p(z)$  never finishes and 
all local costs are positive, each player pays $+ \infty$. 

Let  $|E| = m$.  
The effective cost of a terminal play  $p = p(z)$ 
for player  $i \in I$ 
is the scalar product  $r^i(p) = (u^i, w_p)$ 
of two $m$-vectors: 
$u^i$  is the player  $i$  local cost  $m$-vector  
$(u^i(e) \mid e \in E)$  and 
$w_p$  is the support $m$-vector 
of $(s,t)$-path  $p$, that is, 
$w_p(e) = 1$  if  $e \in p$   and  
$w_p(e) = 0$  if  $e \not\in p$. 

All $n$  players are minimizers.
Thus, a finite $n$-{\em person shortest path (SP) game} is defined. 
We study NS of these games, or more precisely, of the corresponding SP game forms. 

\medskip 

{\bf Shortest path (vector) game forms.} 
Denote by  $\P = \P(s,t)$  
the set of directed $(s,t)$-paths of digraph  $G$.  
The set of outcomes of a shortest path game is 
$\P \cup \{c\}$  
where  $c$  is a special outcome 
merging all infinite plays (lassos). 
Mapping  $g : Z \rightarrow \P \cup \{c\}$,   
defined in the previous subsection, 
is the SP game form.  

\begin{proposition} 
Game form  $g$ is weakly rectangular. 
\end{proposition} 

{\bf Proof.}  
We have two show that each 
$(s,t)$-path  $p \in \P$  fills a box: 

$$g^{-1}(p) = Z^* = Z^*_1 \times \ldots \times Z^*_n 
\subseteq  Z_1 \times \ldots \times Z_n  = Z.$$ 

Suppose that  $g(z') = g(z'') = p$  
for two situations 
$z' = (z'_1, \ldots, z'_n)$  and 
$z'' = (z''_1, \ldots, z''_n)$  in $Z$. 
It is enough to show that 
$g(z) = p$  also for each situation 
$z = (z_1, \ldots, z_n) \in Z$ 
such that  $z_i = z'_i$  or  $z_i = z''_i$  
for every player  $i \in I$. 
By assumption, all $2n$ strategies 
$z'_1, \ldots, z'_n, z''_1, \ldots, z''_n$  
require to follow  $p$  in every position  $v \in p$.
Then obviously, situation  $z$  has the same property 
and, hence, $g(z) = p$  as well. 
\qed 

\medskip

An SP game form  $g$  
is typically not tight, already in case $n=2$.  

Replace in  $g$  every $(s,t)$-path  $p \in \P$  
by its support $m$-vector 
$w_p = (w_p(e) \mid e \in E$), where 
$w_p(e) = 1$  for  $e \in p$  and 
$w_p(e) = 0$  for  $e \not\in p$. 
Furthermore, replace outcome  $c$ in $g$  
by $m$-vector $w_c$  with  $m$  entries equal $+\infty$. 
Then, game form  $g$  will be replaced by a 
(weakly rectangular)  $v^+$-form  $g^+_v$, 
which will be called, an {\em SP $v^+$-form}. 

\medskip 

Return to case  $n=2$. 
The following statement follows immediately 
from definitions.

\begin{theorem} 
\label{t3} 
Every two-person finite SP $v^+$-form is tight 
if and only if bi-shortest path conjecture holds. 
\end{theorem}

Furthermore, by Theorem \ref{t2}, 
tightness and NS are equivalent for $v^+$-forms. 

However, it is an open problem whether tightness holds  
for SP $v^+$-forms and, thus, bi-shortest path conjecture 
remains open too. 

% The Bi-SP conjecture means exactly that all finite two-person SP games 
% (with positive local costs) are NS. 
% Indeed, a situation  $(x, y)$ realizes 
%  a bi-shortest path in $G$  if and only if  $s$ 
% is a  NE in the corresponding two-person SP game. 

\subsection{NS of $n$-person SP $v^+$-forms assigned to symmetric digraphs.} 
{\bf Main result.} For $n > 2$, an $n$-person SP game  
(with positive local costs) may have no NE. 
In other words, the corresponding  $n$-person $v^+$-forms 
are neither NS nor tight. 
An example was constructed for  $n=3$  in \cite{GO14}. 
% Such examples were constructed in \cite{GO15} for $n=4$ 
% and in \cite[Tables 2,3 and Figure 2]{BGMOV18} for  $n=3$.  
However, tightness and NS hold in the following 
important special case. 

Digraph  $G = (V,E)$  is called {\em symmetric} if
each non-terminal move in it is reversible, that is,
$(u,w) \in E$  if and only if  $(w,u) \in E$
unless  $u = t$ or $w = t$.
It was recently shown that every $n$-person SP game 
on a finite symmetric digraph has an NE \cite{BFGV22}.      

\medskip

{\bf Terminal $n$-person games and costs.}
A local cost vector $u : I \times E \rightarrow \RR^m$  
is called terminal if  $u(i, e) = 0$  
for each player  $i \in I$  and 
move  $e \in E$  unless  $e$  is a terminal move. 
Note that these terminal costs 
are arbitrary real numbers: may be positive, negative, or $0$. 
An SP game with terminal costs is called {\em terminal}. 

\begin{remark} 
In this case, it is convenient to replace 
the unique terminal  $t = v_t$  by a terminal set  
$V_T \subseteq V$  assuming that 
each terminal move leads to a separate terminal. 
Then, costs can be defined in terminals rather than for moves.
\end{remark} 

Each (finite) two-person terminal game has a NE, 
yet, this result does not hold for $n$-person games with $n > 2$: 
an NE-free example for $n=4$  was given in \cite{GO15} 
and then, for $n=3$, in \cite{BGMOV18}; 
see Subsection \ref{DGMSGS} for more details. 
% However, in case of symmetric digraphs, NS holds: 
% each $n$-person terminal game on a 
% (finite) symmetric digraph has an NE \cite{BFGV22}.
Yet, the problem is open if we assume 
that the next condition holds: 

\medskip 

(C) Any terminal outcome is better than  $c$  
for each player  $i \in I$. 

\medskip 

Moreover, every known example of an NE-free terminal game 
has the following property:

\smallskip 

(C22) 
There are at least $2$ players each of whom  
has at least $2$ terminals worse than  $c$. 

\smallskip

Is this true for all  NE-free terminal games? 
Such conjecture was called ``Catch 22" in \cite{Gur21A}.  
Also. in this preprint, 
the following strengthening of Catch 22 was suggested: 
 
Partition digraph  $G$  into 
{\em strongly connected components} (SCCs) 
and assign an outcome to each. 
In particular, every terminal vertex is an SCC, 
which will be called {\em terminal}, 
while any other SCC will be called {\em inner}. 
Respectively, the corresponding outcomes 
will be called {\em terminal} and {\em inner}, as well.
See \cite{Gur18} and subsection \ref{DGMSGS} for more details.

Let us merge all inner outcomes 
into one special outcome $c$. 
It is easily seen that 
such operation respects NS and tightness. 
Yet, inverse is not true: 
these properties may appear after merging 
even if they did not hold before. 

\smallskip

Condition (C) and $(C22)$  were generalized 
in \cite{Gur21A} as follows:  

\medskip 

(C$'$) Any terminal outcome is better than 
every inner one for each player  $i \in I$. 

\medskip 

(C$'$22)  % In every NE-free terminal $n$-person game 
There are at least two players each of which 
has at least two terminal outcomes 
that are worse than an inner one.  

\medskip 

In $n$-person terminal case, NS remains open 
if  $n > 2$  and a conditions from 
$\{$(C),(C$'$), (C22), (C$'$22)$\}$ is required.

\subsection{Two versions of the bi-shortest path conjecture}
In the same way we can strengthen 
the Bi-SP conjecture 
assuming that there are several inner outcomes 
each of which is worse than 
every terminal outcome for both players, 
or in other words, that condition (C$'$)  holds. 
% Otherwise, inner outcomes can be ordered arbitrarily. 

\begin{proposition} 
Both versions of the Bi-SP conjecture are equivalent. 
\end{proposition}  

{\bf Proof.} 
Obviously, the strong one implies the standard one. 
Let us show that the inverse implication holds too. 
Consider two cases. 

\smallskip 

Suppose that digraph  $G$  has no $(s,t)$-path. 
Then both versions of the Bi-SP conjecture holds. 
The weak one is trivial, while 
the strong one holds, due to NS result of \cite{Gur18}. 

\smallskip 

Suppose that digraph  $G$  has an $(s,t)$-path   
and the Bi-SP conjecture holds. 
Then, the corresponding $v^+$-form $g^+_v$ is NS, 
that is, for any cost functions 
$u^A$  and  $u^B$, 
game  $(g^+_v, u^A, u^B)$  has a NE  $(x,y)$  
such that  $g_v+(x,y)$  is a terminal vector. 
Let us merge all inner outcomes. 
Then, $(x,y)$  remains a NE, by condition  (C$'$). 
\qed

\medskip 

If digraph  $G$ is symmetric then 
there exists a unique inner outcome and, hence,  
both version of the Bi-SP conjecture coincide. 
Moreover, for symmetric digraphs NS 
holds even for the $n$-person SP game forms \cite{BFGV22}. 
In contrast, without assumption of symmetry,  
already the weak version fails for  $n=3$  \cite{GO14}, 
and for  $n=2$  both versions of the Bi-SP  conjectures 
are equivalent and open. 

\medskip 

For terminal costs, 
the strong version is proven for $n=2$ in \cite{Gur18}, 
while for  $n>2$ even the weak version fails; 
the counterexamples were given    
for  $n=4$  in \cite{GO15} and 
for  $n=3$  in \cite{BGMOV18}. 
Yet, the weak and strong versions are open 
if  we assume  (C)  and  (C$'$), respectively,  
and remain open if we weaken conditions (C)  and  (C$'$) 
requiring instead their Catch 22 versions, (C22)  and  (C$'$22). 

\bigskip 

{\bf Acknowledgements}
The paper was
prepared within the framework 
of the HSE University Basic Research Program.
% Moscow, Russia.
% and funded by the RSF grant  20-11-20203.
% The author is thankful to Endre Boros for many helpful remarks.

\end{document}